\documentclass[final]{SolarPhysics}
\usepackage{graphicx}
\usepackage{geometry}
\usepackage{natbib}
\usepackage[english]{babel}
\usepackage{url}
\usepackage{txfonts}

\newcommand{\arcsec}{\hbox{$^{\prime\prime}$}}
\def\pasp{{\it Pub. Astron. Soc. Pac.}}%  % Publications of the ASP
\def\solphys{{\it Solar~Phys.}}%           % Solar Physics 
\def\aj{{\it Astron. J.}}%
\def\apj{{\it Astrophys. J.}}%          % Astrophysical Journal
\def\ao{{\it Appl.~Opt.}}%          % Applied Optics
\def\aap{{\it Astron. Astrophys.}}%          % Astronomy and Astrophysics
\newcommand\newblock{\hskip .11em\@plus.33em\@minus.07em}
\newcommand{\degree}{\ensuremath{^\circ}}

\begin{document}
\begin{article}

\runningtitle{ATMOSPHERIC DISPERSION AND HIGH-RESOLUTION SOLAR SPECTROSCOPY}
\runningauthor{K. REARDON}

%\begin{center}
\begin{opening}

%\centering
\title{THE EFFECTS OF ATMOSPHERIC DISPERSION 
        ON HIGH-RESOLUTION SOLAR SPECTROSCOPY}
\author{KEVIN P. \surname{REARDON}}

\institute{INAF\,/\,Osservatorio Astrofisico di Arcetri, 
Largo E. Fermi 5, Florence, 50125, Italy\\
(e-mail: kreardon@arcetri.astro.it)}

\date{\received{Recevied: 02 August 2006}}
%\date{02 August 2006} 
%\accepted{22 September 2006}
\date{(Received 02 August 2006 ; Accepted 22 September 2006)}

\begin{abstract}
We investigate the effects of atmospheric dispersion on observations 
of the Sun at the ever-higher spatial resolutions afforded by 
increased apertures and improved techniques. 
The problems induced by atmospheric refraction are particularly
significant for solar physics because the Sun is often best observed
at low elevations, and the effect of the image displacement is not merely a loss
of efficiency, but the mixing of information originating from different
points on the solar surface.
We calculate the magnitude of the atmospheric dispersion for the 
Sun during the year and examine the problems produced by this 
dispersion in both spectrographic and filter observations.
We describe an observing technique for scanning
spectrograph observations that minimizes the effects of the atmospheric dispersion
while maintaining a regular scanning geometry. Such an approach could be useful
for the new class of high-resolution solar spectrographs, such as SPINOR, 
POLIS, TRIPPEL, and ViSP.
\end{abstract}

%\maketitle
\end{opening}
%\end{center}

\section{Introduction}
Solar physics increasingly relies on high-spatial-resolution, high-spectral-purity 
observations in order to determine the three-dimensional structure of the
solar atmosphere. In such observations multiple spectral lines are often used because
they sample different heights or are sensitive to different parameters in the
solar atmosphere \citep*[\textit{e.g.} ][]{2005A&A...439..687C}. In particular, the comparison of
polarization signals from different lines, often using sophisticated inversion
techniques, allows a more accurate determination of the magnetic field and
physical conditions at the source \citep*{2000ApJ...535..475B}.

However, in order for the comparison or correlations among different spectral lines to be
useful, it is often necessary that the lines be observed not only cospatially,
but also within the evolutionary timescales of the resolved
atmospheric element. To assure that these conditions are met, it is
important to take into consideration the effects of atmospheric dispersion, in
particular for slit-based spectrographs \citep{1966AJ.....71..190S}.

The magnitude of the atmospheric refraction varies with zenith distance, an
effect called ``spatial differential refraction''. The variation with
wavelength of the index of refraction of air causes an additional variation, 
called  ``spectral differential refraction''
or ``atmospheric dispersion''. This produces an offset in the direction
perpendicular to the horizon of the apparent position of the same object
viewed at different wavelengths. This is of particular importance for
spectrographs because if the slit is not oriented along the direction of this
dispersion, different positions on the sky will be sampled at different
wavelengths.

For nighttime observations the effects of spatial differential refraction and
atmospheric dispersion have been studied in the past, in particular by
\cite{1982PASP...94..715F}, which led to greater care being taken in
the slit orientation by nighttime observers. New multi-object spectroscopic
techniques brought a slightly different set of problems as discussed by
\citet{1988PASP..100.1582C}, \citet{1989PASP..101.1046D}, and most recently by
\citet{2005A&A...443..703S}. These authors discuss the atmospheric
effects on various types of multi-object spectrographs in different observing
regimes and how to optimize the observing setup in the presence of these
effects.

However, solar observations present some characteristics that are significantly
different from nighttime observations, making it difficult to apply the
results of these previous works to the solar case. Firstly, because the
insolation by the Sun produces local heating that results in atmospheric
turbulence close to the telescope itself, high-resolution solar observations are
often best performed when the Sun is at a relatively low elevation. Indeed, at
many sites, the most stable local atmospheric conditions are obtained in the few
hours after sunrise when the Sun is at elevations as low as 10\degree~\citep{Hill2004.sitesurvey, SocasNavarro2005.sitesurvey}. 
This constraint on atmospheric stability means that the
observations cannot necessarily be optimized, for example by choosing the
optimal hour angle at which to observe a given object, in order to reduce 
the effects of the atmospheric refraction \citep{1989PASP..101.1046D}. 
Further, for nighttime observations of a ``bright'' object against a darker
background, the principal effect of atmospheric dispersion is a reduced system
efficiency at wavelengths where the image of the object is shifted off of the
slit (or other entrance aperture). In solar observations, such shifts will result
instead in illumination from different elements of the solar atmosphere at
different wavelengths, which can lead to difficulties in the physical
interpretation of the data. In addition, solar telescopes are generally 
not fitted with atmospheric dispersion correctors. One example to the 
contrary is the Swedish one-meter Solar Telescope, which was designed 
with the possibility to compensate for the atmospheric dispersion as a 
side benefit of the Schupmann system used to correct the chromatic 
aberration of the telescope's singlet objective \citep{1999ASPC..183..157S}.

Current solar telescopes are able to resolve features as small as 0.1'' 
on the solar surface \citep{2005A&A...435..327R}. while future telescopes, 
such as GREGOR \citep{Volkmer2005.GREGOR} and the
Advanced Technology Solar Telescope \citep[ATST; ][]{Keil2004.ATST}, will have even greater 
resolutions, the latter with a diffraction limit as small as 0.025''. 
Advances in multi-conjugate
adaptive optics and the maturing of image-reconstruction techniques are making
it increasingly routine to achieve diffraction-limited images over large fields
of view. It should be noted that correlation tracking or adaptive optics
generally provide no remedy to the problem of spectral differential refraction. 
These systems correctly stabilize the solar image at their operating wavelength 
but do nothing to correct the relative offsets due to the atmospheric dispersion 
of images at other wavelengths.

As the diffraction limit decreases, the atmospheric dispersion, which is 
independent of aperture, becomes increasingly significant with respect to the
resolution element and deserves accurate consideration. In Section~\ref{sec:refraction} we
calculate the annual and diurnal variation of the atmospheric dispersion 
for the Sun for a specific location. In Section 3 we examine the effects of
atmospheric dispersion on scanning-slit spectroscopy and describe an observing
procedure that minimizes the effects of this dispersion while also maintaining
regular spatial sampling. Finally, in Section 4 we discuss the effects of 
atmospheric dispersion for filter-based images.

\section{Atmospheric Refraction}\label{sec:refraction}
\subsection{Refraction Calculations}

The formula for the index of refraction of moist air (\emph{n}) has been most
recently described by \citet{1996ApOpt..35.1566C, 1999ApOpt..38.1663C}, who presents updated versions
of Edlen's \citeyearpar{1966Metro...2...71E} equations for calculating the index
of refraction based on the temperature, humidity, pressure, and CO$_{2}$
concentration of the atmosphere. Atmospheric refraction will produce a
significant change in the apparent zenith angle that varies with index of
refraction and the altitude of the observed object. The magnitude of the
refraction can be approximated by \citep{1966QB145.W83......}

\begin{equation}
\label{Eqn:r}
R=r\ (1-\beta)\tan z_{a}-r\ (\beta-\frac{r}{2})\tan^{3}z_{a}
\end{equation}

\noindent where $r=n - 1$ is the refractivity at the observation site, $z_{a}$ is
the true zenith angle, and $\beta=H_{0}/r_{0}$ is the ratio of the height of the
equivalent homogeneous atmosphere at the observatory to the 
geocentric distance of the observatory. The value of $H_{0}$ is approximately
8 km and $\beta$ can be approximated as 
$0.001254\left(\frac{T_{0}}{273.15}\right)$, 
where $T_{0}$ is the temperature in Kelvin at the observatory \citep{1996PASP..108.1051S}.
This formula for the refraction is accurate to better than
approximately 1\arcsec for zenith angles less than 75\degree. More accurate
determination of the absolute refraction for the correction of
spatial differential refraction requires a tropospheric lapse-rate term
\citep{2005ASPC..338..134C} or a full integration over the atmospheric path
\citep{Auer.Standish2000}. 
Since the fields of view utilized in high-resolution solar physics are generally
only a few arcminutes across, the distortions in the field produced by spatial
differential refraction are usually less than a few arcseconds. This will 
cause time-dependent image distortions in observations covering a
large range of zenith angles, but absolute positions are generally not 
required for solar data analysis.

We are instead concerned with the dependence of the index of refraction on 
wavelength which causes proportional
variations in $R$, called atmospheric dispersion. For zenith angles up
to approximately 70\degree, the dispersion can be approximated using only the
first term in Equation~(\ref{Eqn:r}). For slightly larger zenith angles, 
considering that
1 - $\beta$ is very close to unity and that in most situations $\beta$ is at least
an order of magnitude greater than $r/2$, the dispersion can be approximated
using

\begin{equation}
\label{Eqn:deltar} % is used to refer this table in the text
\Delta R=\left(r_{\lambda}-r_{\lambda_{0}}\right)\left(\tan z_{a}+\beta\tan^{3}z_{a}\right)
\end{equation}

\noindent where $\lambda$ and $\lambda_{0}$ are the observation and reference
wavelengths. We have also calculated the magnitude of the dispersion
using numerical integration through a model atmosphere following the
technique outlined by \citet{Seidelmann.1992.expsup} and observe that the above
equation reproduces the atmospheric dispersion to better than 0.05'' 
up to zenith angles of 80\degree. Closer to the horizon this
linear approximation rapidly breaks down.
\citet{2005A&A...443..703S} examined the dependence
of the index of refraction on the input parameters and points out
that the magnitude of the atmospheric dispersion is most sensitive
to variations in the atmospheric temperature and pressure, indicating
the importance in knowing these values for the accurate calculation
of the atmospheric dispersion for a given observation.

The direction of the shift between images at different wavelengths
always remains along the vertical circle, that is perpendicular to 
the horizon. However, this direction rotates continuously
during the day with respect to the axes of the celestial coordinate system. 
The angle ($\eta$) between the vertical circle and the 
hour circle passing through the 
celestial poles and the observed object is called 
the \textit{parallactic angle}. Away from the
Earth's poles, the parallactic angle will be zero only when the observed
object is on the meridian. \citet{1982PASP...94..715F} gives the
following equation for parallactic angle:

\begin{equation}
\label{Eqn:parang} % is used to refer this table in the text
\sin\eta=\sin h\times\sin(\frac{\pi}{2}-l)/\sin z_{a}
\end{equation}

\noindent where $h$ is the object's hour angle (positive west of meridian)
and $l$ is observer's latitude.

\subsection{Atmospheric Dispersion for the Sun}

\begin{table*}
%\begin{center}
%\centering % used for centering table 
\caption{Selected wavelengths and their calculated indices of refraction} % title of Table
\label{Table:1} % is used to refer this table in the text
\begin{tabular}{c c c}
\hline 
Wavelength (nm) & Index of Refraction & Relevant Solar Spectral Lines\tabularnewline
\hline
388& 1.00020152 & CN Band \\ % \tabularnewline
400& 1.00020103 & Ca\,{\footnotesize II H\,\&\,K} \\ % \tabularnewline
430& 1.00020001 & CH Band \\ % \tabularnewline
630 & 1.00019662 & Fe\,{\footnotesize I} \\ % \tabularnewline
850 & 1.00019534 & Ca\,{\footnotesize II} Triplet \\ % \tabularnewline
1080 & 1.00019476 & He\,{\footnotesize I}, Fe\,{\footnotesize XIII} \\ % \tabularnewline
1600 & 1.00019424 & Fe\,{\footnotesize I} \\ % \tabularnewline
\end{tabular}
%\end{center}  
\end{table*}

\begin{figure*}
\centering
\includegraphics[bb=30 0 480 290,width=14cm,height=9cm]{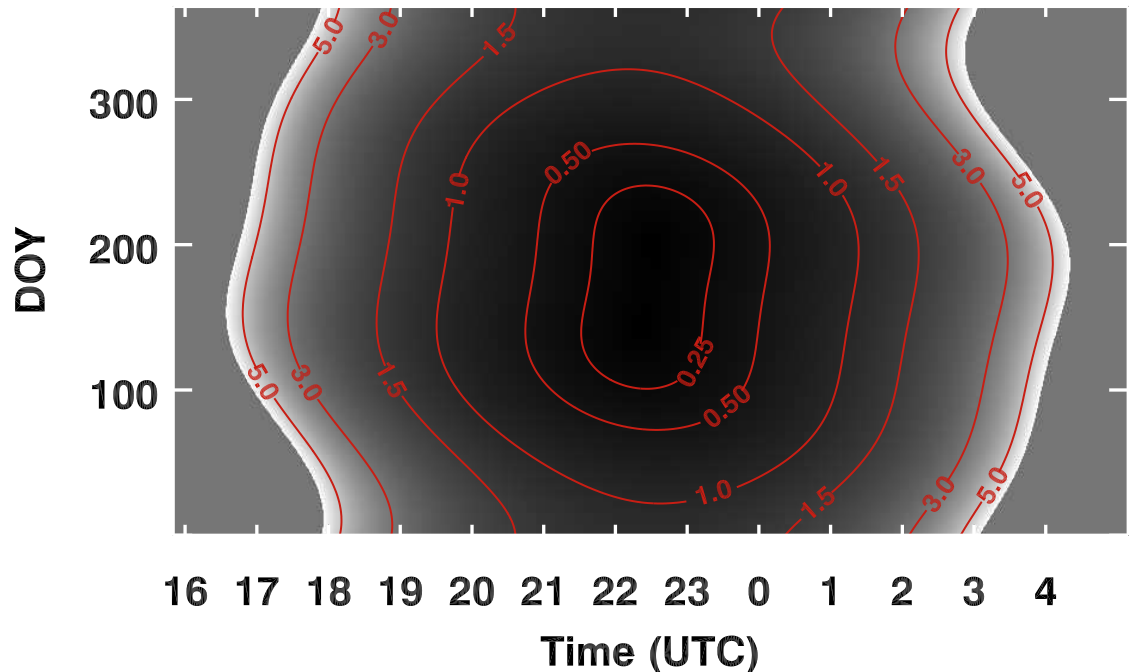}
\caption{Magnitude of the atmospheric
dispersion, in arcseconds, at Haleakal\=a, Hawai`i for the Sun between
wavelengths of 400 and 850 nm. For each day of the year, shown on
the vertical axis, the dispersion is shown for all times when the
true elevation of the Sun is greater than 10\degree.}
\label{Figure:1}
\end{figure*}

We use the relationships defined above to estimate the magnitude and direction
of the atmospheric dispersion expected for observations from Haleakal\=a,
Hawai`i. We chose this site since it is the proposed site of the new four-meter
solar telescope ATST and high-resolution, multi-wavelength observations have
been indicated as important goals for this new facility. Similar calculations
for La Palma in the Canary Islands, another prime location for high-resolution
solar observations, are shown in Appendix \ref{appendixa}. First, the right
ascension and declination of the Sun are calculated with a one-minute time step
for each day throughout the year (the calculations were specifically done for
2006, but are applicable to any year). Then using the coordinates of
Haleakal\=a (latitude: 20.71\degree; longitude: -156.25\degree; altitude: 3055 m), the azimuth
and altitude of the solar disk center was calculated for each minute. 
Using the equations given by Ciddor (1996) we then calculate the
atmospheric index of refraction at several different typical wavelengths of
importance in solar observations, shown in Table 1. We use typical conditions
for Haleakal\=a (Temp: 11 $^{\circ}$C; Pressure: 71\,000 Pa; Humidity : 30\%; 
CO$_{2}$ mixing ratio: 380 ppm) as
measured at the Mees Solar Observatory \citep{mees_weather}, 
except for the CO$_{2}$ fraction, which is the value measured on Mauna Loa. 
We combine the indices of refraction
for two different wavelengths with the zenith distance of the Sun to calculate
the magnitude of atmospheric dispersion during the course of the year as shown
in Equation (\ref{Eqn:r}). The programs for calculating the refractivity and
refraction are available in SolarSoft 
({\small \url{http://www.lmsal.com/solarsoft/}}) and at 
{\small \url{http://www.arcetri.astro.it/science/solar/}}, as well as Solar Physics 
Electronic Supplementary Material.

We show the calculations for the magnitude of the atmospheric dispersion between
400 and 850 nm in Figure \ref{Figure:1}. We choose these two sample wavelengths
since they represent a fairly typical observing combination and cover
approximately the full visible spectral range. Since Equation (\ref{Eqn:deltar}) shows
that the atmospheric dispersion scales almost linearly with the difference 
of the indices of refraction for zenith distances less than approximately 
80\degree, it should be straightforward to apply the following
discussions to other wavelength combinations. The figure shows the calculated
value only for those times when the Sun is more than 10\degree~above the horizon.
Figure \ref{Figure:1} shows that even at moderate spatial resolutions ($\sim$1\arcsec) 
the atmospheric dispersion remains significant for the first few
hours after sunrise or before sunset throughout the course of
the year. At higher resolutions, the dispersion will need to be taken into
account at almost all times. 
 
The shift between images obtained at different wavelengths due to atmospheric
dispersion could be removed by aligning on common solar structures, such as the
granulation pattern, taking care not to be 
biased by variations in the structures observed at different wavelengths. 
An alternate method would be to apply a correction based on the calculated 
magnitude and direction of the atmospheric dispersion given the local 
meteorological conditions. It may be desirable to calculate the 
atmospheric refraction using more accurate models or through numerical 
integration of a standard atmosphere \citep{2004AJ....127.3622Y}.
A combined approach could also be developed using 
the measured offset between images at two suitably selected wavelengths 
(\textit{i.e.} well separated and observing similar structures) and interpolating 
the offset to other wavelengths based of the relative variation of the 
index of refraction with wavelength. Since the magnitude
of the atmospheric dispersion varies with the zenith angle and with variations 
in the local atmospheric conditions, any alignment of multi-wavelength 
observations obtained over an extended period of time should  
take into account the temporal variations of the 
atmospheric dispersion.

\section{Atmospheric Dispersion and Spectrographic Observations}

Classical long-slit spectrograph observations are widely used in solar
physics to record an approximately one-dimensional slice of the solar
atmosphere often in multiple spectral lines covering a significant wavelength
range. For many scientific questions it is necessary to obtain spectral
information over a fully filled 2-D field with reasonably high time
resolution. In order to achieve this, the field of view is stepped
across the spectrograph slit in a direction perpendicular to the 
orientation of the slit, recording
separate spectra at each position. There is a new group of multi-wavelength
(various ranges from 390 -- 1600 $\mu$m), high-resolution ($\leq$
0.3 \arcsec/pixel), scanning spectrographs currently being used
in solar physics, including 
SPINOR \citep{2006.SocasNavarro.SPINOR},
POLIS \citep{2005A&A...437.1159B}, and 
TRIPPEL \citep{trippel_web}, or being constructed, such as the 
ViSP for the ATST \citep{2005SPIE.5901...60E}. Due to the limitations
given by the photon flux, it is often necessary with these instruments
to integrate for several seconds or more at each slit position. A
single scan of a sizable area of the solar surface may require tens
of minutes or up to an hour. In addition, since the temporal evolution of
the solar structures or the accurate measurement of oscillatory behavior
is important, it is often necessary to track and repeatedly scan the
same region in the solar atmosphere continuously for periods of several
hours or more.

\subsection{Observational Effects}

The spectrograph slit is placed in a focal plane, extracting a portion
of the image formed on the slit. However, the atmospheric dispersion will cause 
images at multiple wavelengths to be formed at different positions on the 
focal plane with respect to the slit. If a significant component of the 
separation between the images at different wavelengths lies perpendicular 
to the length of the slit, then the spectrograph will sample different 
regions on the solar atmosphere at each wavelength. 
By orienting the slit along the direction of the chromatic separation, 
the same region will be observed (with a slight shift along the length of 
the slit) at all wavelengths. Thus the spectra at different wavelengths can
subsequently be aligned using the techniques described above. This is the
approach taken, for example, by \citet{1972SoPh...25...81B},
\citet{1981SoPh...71..237C}, and more recently by
\citet{1994ASPC...68..389B} and \citet{2004A&A...420.1141R}.

\begin{figure*}
\centering
\includegraphics[bb=30 0 480 290,width=14cm,height=9cm]{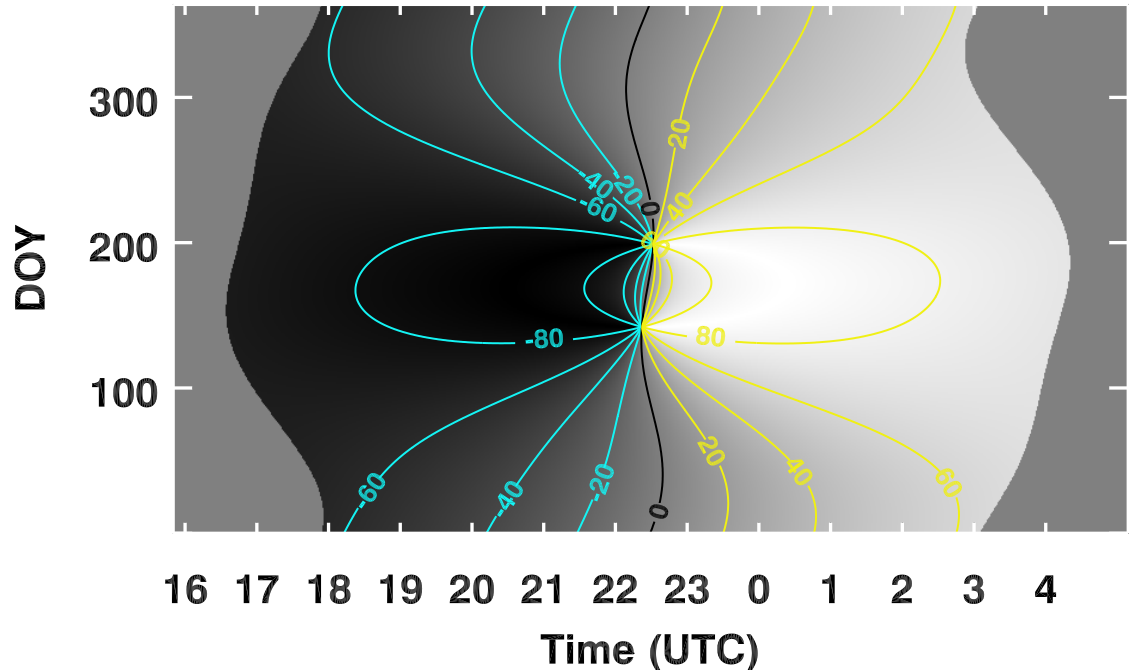}
\caption{The parallactic angle in degrees at Haleakal\=a,
Hawai`i for the Sun for all times when the true elevation of the Sun
is greater than 10\degree. Negative angles are shown in blue, positive
angles in yellow.}
\label{Figure:2}
\end{figure*}

Using a scanning spectrograph with the slit held fixed along the direction
of the atmospheric dispersion thus allows observations for which at each 
slit step the same portion of the solar surface is observed at
all wavelengths. However, this approach has generally been avoided 
because in this case the slit undergoes 
a constant rotation with respect to the celestial object being observed. 
This can be seen in Figure \ref{Figure:2} where the parallactic angle 
has been calculated from Equation (\ref{Eqn:parang}) using the same ephemeris 
employed for Figure \ref{Figure:1}. It can be seen that during the 
Winter the parallactic angle undergoes a continuous variation during 
the day, while in the Summer when the Sun passes nearly overhead at the 
latitude of Haleakal\=a, the parallactic angle remains nearly constant 
except for a rapid variation near local noon corresponding to the large 
changes in azimuth as the Sun passes near the zenith.

\begin{figure*}
%\resizebox{\hsize}{!}
\centering
\includegraphics[width=7cm,height=8cm]{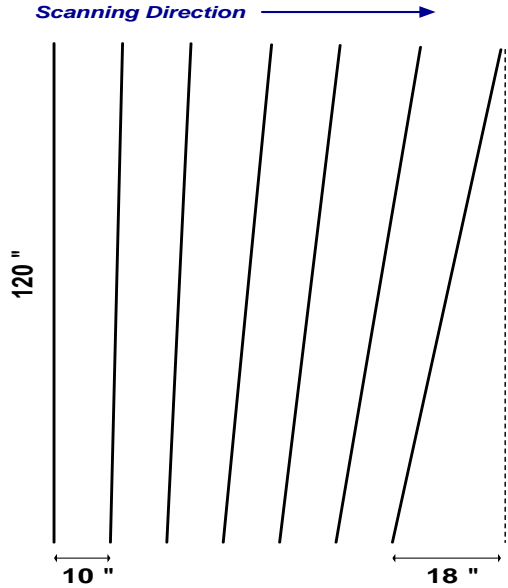}
%{\includegraphics{diff.rerf.spect.scan.eps}}
\caption{The alignment of the slit on a celestial coordinate system, 
showing the effects of the rotation
during an example spectral scan when the slit is held fixed at the parallactic
angle. The slit is 120\arcsec long, steps at a rate of 
20 steps per arcsecond, and requires 3 seconds per step (requiring 
1 minute to scan 1\arcsec).
The slit position at intervals of 10 minutes or 10\arcsec is shown.
This 60\arcsec scan would take one hour, and in that time the image
has rotated 8.5\degree, resulting in an 18\arcsec deviation in
the slit position from the starting orientation.}
\label{Figure:3}
\end{figure*}

This apparent rotation of the slit orientation will produce complications 
in the geometry of the spatial scan, as illustrated in Figure \ref{Figure:3}. 
The extent of this distortion will depend on the rate of change of 
the parallactic angle and the amount of time it takes complete the
scan. Subsequent scans of the same area, taken with the Sun at the
different altitude and hour angle, will have differing amounts of 
distortion and will all need to be mapped to a common grid
in order to be compared. It should also be noted that this rotation 
of the observed object with respect to the direction of the atmospheric 
dispersion implies that multi-wavelength fixed-position 
(\textit{i.e.} with no spatial scanning) spectrographic observations are inherently 
plagued by the fact that
there is no means to keep a one-dimensional slice of the solar surface 
observed at multiple wavelengths fixed with respect to a spectrograph slit 
over time. 

Generally then, when building up a two-dimensional field with a 
scanning spectrograph the slit is held fixed with respect to the
celestial coordinate system and thus provides a regular sampling of 
the solar surface. In this case there will usually be some component
of the atmospheric dispersion which is perpendicular to the slit.
Since a full map of the solar surface is observed, it is possible 
to correct for an arbitrary direction
of the atmospheric dispersion by applying the appropriate shifts 
to the data cubes in directions both parallel and perpendicular to
the slit \citep*{1997ApJ...477..485S, 1999ApJ...517.1013L,
2003A&A...412..541M}. In most cases the changes in the
magnitude and the direction of the atmospheric dispersion during a 
single scan are not significant and the mean value can be applied, 
although at higher resolutions the variations during a scan may become
important. However, even though the shifts between maps at different
wavelengths can be removed, there remains a problem in that the
spectra of the same portion of the solar surface may be obtained at 
different times in different wavelengths. The time delay 
between sampling the same
position at two different wavelengths can be given by

\begin{equation}
\label{Eqn:deltat} % formula for time delay in spectral scanning
\Delta t=\frac{\Delta R{_{\perp}}}{s} \times t{_{step}}
\end{equation}

\noindent where $\Delta R{_{\perp}}$ is the magnitude of the
chromatic separation perpendicular to the slit, $s$ is the size
of the scan step, and $t{_{step}}$ is the time required
for each step. Since the temporal evolution can be rapid with respect
to the scanning speed, the correlations between measurements at 
multiple wavelengths could be compromised by changes in the 
solar structures or differing phases of solar oscillations.

Consider, for example, an observation using a scan step of 0.2\arcsec 
(corresponding to 140 km on the solar surface) and a four-second exposure
time per scan position. The evolutionary time scale of a 140 km 
element in the solar photosphere is on the order of 20 seconds or more
(chromospheric timescales will be even shorter). Observations requiring 
direct comparison between structures or spectral profiles measured at 
different wavelengths should all be acquired within this time span.
If the magnitude of the chromatic separation perpendicular
to the slit is greater than approximately 1\arcsec, this condition
will not be met and comparisons among multiple wavelengths,
even after spatial alignment to remove the offsets produced by the
dispersion, will remain problematic. Examination of Figure 
\ref{Figure:1} shows that the magnitude of the dispersion is 
greater than 1\arcsec during the first three hours after 
sunrise or before sunset. As even higher
resolutions are achieved, the relative importance of the atmospheric
dispersion will increase with respect to the slit width and scan step.

\begin{figure*}
\centering
\includegraphics[bb=30 0 480 300,width=14cm,height=9cm]{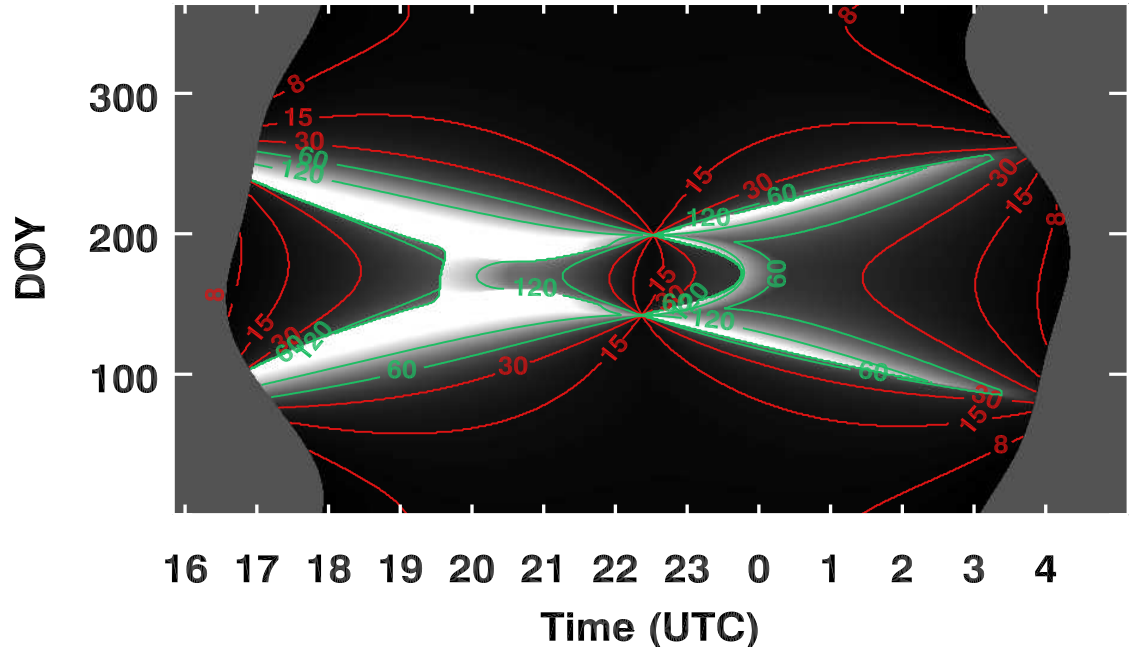}
\caption{The duration in minutes for which spectra
obtained at two different wavelengths will remain aligned within one slit
width. The slit is initially oriented perpendicular to 
the horizon but the image is rotated during the scan to maintain a constant 
orientation with respect to the celestial coordinates.
The figure is calculated for observations obtained at Haleakal\=a at 400 and 850
nm and with a slit width of 0.05\arcsec.}
\label{Figure:5}
\end{figure*}

\subsection{Stepped Parallactic Angle Alignment}

While the bulk rotation of the celestial coordinate frame with respect
to the direction of the atmospheric dispersion cannot be eliminated, 
it is possible to devise a method that would at least allow for the 
regular sampling of a 2-D area of the solar surface while ensuring that 
the same elements on the solar surface are observed simultaneously at multiple 
wavelengths. This approach takes
advantage of the fact that the rotation between the image plane and
the slit operates over the full slit length, while for the shift between
images at different wavelengths the lever arm is only as long as the
chromatic separation between the images. This gives a greater tolerance in the
orientation of the slit so that it can be allowed to rotate away 
from the true vertical direction within certain limits, allowing the slit to
be maintained at a fixed orientation in celestial coordinates.

The basic approach is that prior to performing a spatial scan with the spectrograph,
the slit is oriented at an angle corresponding to the mean parallactic angle for
the full scan to be performed. During the scan, the solar image is rotated such
that its orientation is held fixed with respect to the spectrograph slit,
resulting in rectilinear sampling of the observed field. Eventually the
parallactic angle will change such that the images at the wavelengths being
observed will be significantly offset perpendicularly to the slit. This will
require that the slit be oriented along the new mean parallactic
angle before resuming the image rotation necessary to maintain the slit at a fixed
direction on the solar surface. When the parallactic angle is stepped all
subsequent scans will have a different overall orientation in celestial
coordinates, but this can be dealt with through a simple bulk rotation of the
datacube to a common orientation.

The metric of interest in this case is the amount of time that the
slit can be held at a fixed orientation in celestial coordinates before
the change of the parallactic angle causes the shifts perpendicular
to the slit for images at different wavelengths to be significant.
A perpendicular shift of half of the slit width can be considered
significant since it will cause the primary contribution to the flux
at different wavelengths will come from different spatial positions.
Since the half of a slit width shift is acceptable in either of the
two directions perpendicular to the slit, we take one slit width as
the cutoff for the maximum allowable shift.

The magnitude of the perpendicular shift at any moment depends on the difference
between the starting ($\eta_{0})$ and actual ($\eta)$ parallactic
angle and the magnitude of the atmospheric dispersion at that moment.
Taking each one-minute time step as the starting point, we calculate
the value of $\sin\left(\eta-\eta_{0}\right)\Delta R$ for all subsequent
times during that day. We then determine the first moment when this
function exceeds the defined cutoff of one slit
width. The result is a measure of how long, starting from the slit
oriented along the direction of the chromatic separation, the image can
be rotated to maintain the slit at a fixed orientation in 
celestial coordinates without introducing significant offsets 
transverse to the slit at different observed wavelengths.

Figure \ref{Figure:5} shows the allowable durations for all times during the year,
calculated for an example observation with the ATST spanning the wavelengths
400 and 850 nm, and with a slit width of 0.05\arcsec. Observations at
this resolution will require excellent seeing and adaptive 
optics stabilization, but obtaining multiwavelength spectral information
on solar structures at this scale is an important science driver for ATST.
It can be seen that the calculated duration can vary strongly 
during the day and shows different behaviors at different times 
of the year. In the winter
months from October through March, the allowable time is often 15
minutes or less. Considering a typical exposure time of approximately
five seconds, this only allows for 180 step positions, which may allow a
scan of less than 10\arcsec on the solar surface. There are
also two one-month periods, starting in April and August, when the
slit can be held at a fixed orientation with respect to the celestial
coordinates for an hour or more. These periods might be best
employed for certain observations requiring spatially and temporally
extended observations at multiple wavelengths.

A further optimization in the calculation of the allowable durations 
is possible by not forcing the slit to be aligned strictly
in the direction of the atmospheric dispersion
at the start of the observation, but rather to permit it to be set
to any angle, while still maintaining the perpendicular displacement
to be less than the cutoff value. In some cases this can allow for
longer periods of observations without the need to adjust the orientation
of the slit with respect to the vertical direction. The gain is seen
mostly at midday when the Sun passes near the zenith and the parallactic
angle rotates rapidly but $\Delta R$ is very small.

\section{Filter Observations}

\begin{figure*}
\centering
\includegraphics[bb=20 20 380 320,width=12cm,height=7cm]{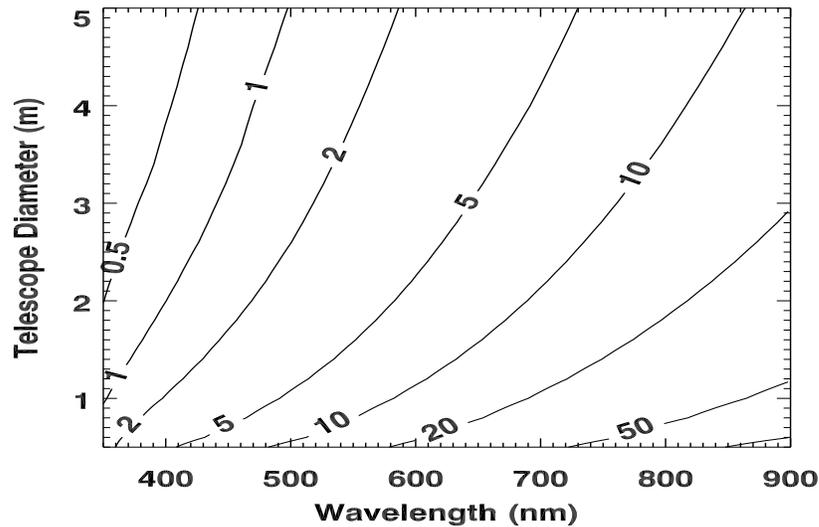}
\caption{The FWHM in nanometers of a simple two-cavity filter for which the smearing 
caused by the relative atmospheric dispersion within the filter passband
will reduce the peak intensity of the Airy profile of the 
telescope with a given aperture by a factor of 0.8.}
\label{Figure:6}
\end{figure*}

We have been primarily concerned with the effect of atmospheric dispersion
on spectrographic observations of the Sun, but obviously the same
effect will present in filter observations as well. While in this latter
case the acquisition of instantaneous two-dimensional maps at each 
wavelength allow for the application of arbitrary shifts to align 
the images, it should be remembered that the appropriate 
shifts to coalign different wavelengths will vary in magnitude 
and direction during the course of the day.

One concern with imaging through a filter is the smearing caused by
the separation between the solar scene at the different wavelengths 
transmitted by the filter. The significance of this smearing depends
on the magnitude of the atmospheric dispersion, the width of the 
filter passband and the resolution achieved in the final image. 
In order to quantify the effects of this smearing, we estimate the 
change in the diffraction limited PSF for a telescope with a 
given aperture.

The transmission profile for a filter provides the percentage of the
incident flux transmitted at each wavelength. For each transmitted wavelength,
we can calculate, for a specific set of atmospheric conditions
and observing circumstances, the associated atmospheric dispersion. For each 
spectral position in the profile then we can calculate the offsets relative
to the central wavelength by setting $\lambda_{0}$ in Equation (\ref{Eqn:deltar}) 
to the central filter wavelength. By combining the calculated offsets with 
the transmission profile, we construct a weighted distribution of 
shifts in the resulting image. This spatial smearing profile can
be convolved with the diffraction limited PSF for a telescope of a
given aperture to determine how the telescope PSF is broadened by 
the effects of the atmospheric dispersion. This smearing will be 
produced only along the direction of the 
atmospheric dispersion.

We perform this calculation for a range of wavelengths and telescope
apertures. We calculate the refractivity for the atmospheric conditions 
given in Section 2.2 and for observations at an elevation of 15\degree~
above the horizon. We calculate the transmission profile for a 
ideal two-cavity filter centered on each wavelength in a range 
from 350 to 900 nm. We then convert the transmission profile into 
a spatial smearing profile that is convolved with the Airy 
function for telescopes with apertures ranging from 0.5 to 5 meters. 
For each combination of wavelength and aperture, we find the FWHM 
of the filter profile that results in a reduction of the peak 
transmission of Airy profile by a factor of 0.8. This value 
for the Strehl ratio implies a possibly tolerable but not 
altogether negligible degradation in the image quality. This 
smearing should be included during the design process in the 
overall error budget in determining the final image quality 
for a given instrument

The results of this calculation are shown in Figure \ref{Figure:6}, where the
contours indicate the FWHM of the two-cavity filter that achieves the
defined Strehl ratio. Even at the current apertures of one meter or 
less, typical filters with a full width of ten nm can result in 
a notable image degradation at shorter wavelengths. With a four-meter 
class telescope this constraint becomes more limiting, reducing the 
usable filter passbands and offsetting the gain in photon flux with
the larger aperture. For example, G-band observations with a four-meter
telescope may be limited to a 0.5 nm passband, down from the one nm
filters currently in common use.

\section{Discussion}

We have evaluated the magnitude and direction of the atmospheric dispersion
for the Sun for all times during the year. The relative offsets between
images at different wavelengths are significant, especially in the
low-elevation observations typical for high-resolution solar observations.
The magnitude of this effect changes during the course of the day,
which means that the alignment between images at different
wavelengths will be a function of the time of observations. Image alignment
cannot rely solely on reference points within the telescope (grids,
crosshairs, \textit{etc.}), but must be either measured from the solar structures
or calculated using the known functions for atmospheric refraction 
(or some combination of the two).

Spectroscopic observations covering a broad range of wavelengths and
requiring a direct comparison among different spectral lines, should
be made by aligning the slit along the parallactic angle, although this
will cause a rotation of the slit in celestial coordinates and
a distortion of the scanning geometry. This will require a separate
remapping of all of the data to a common grid, but it has the advantage 
for telescopes on an alt-azimuth mount that no optical image
derotator is required for some instrument positions and the slit is 
held at a fixed position with respect to the telescope, possibly 
simplifying the polarization calibration. Also, it is simpler to 
maintain the slit at a fixed orientation with respect to the horizon, 
since the spatial differential refraction will produce an small 
but non-negligible extraneous rotation of the observed celestial 
object, in addition to the rotation caused by the changing 
parallactic angle, that will need to be calculated and corrected.

We have described a method allowing a regular sampling in celestial
coordinates that allows the slit to deviate from the parallactic angle
as long as the perpendicular shifts do not exceed a defined
limit. Using this limit we can calculate the acceptable amount
of time that the solar image can be rotated to maintain a fixed orientation
to the celestial coordinates without causing an unacceptable shift
of images at different wavelengths perpendicular to the slit direction.
The calculated period obviously depends on the slit width being used
and the wavelength separation, but it still provides an observational 
constraint even for existing telescopes and resolutions, 
as can be seen in Appendix A.

The proper consideration of the atmospheric dispersion places constraints
on the observational configuration. If the orientation of the slit is dictated by
the parallactic angle then it cannot be oriented based on the solar
structure to be observed. For example, the slit can only be placed parallel or
perpendicular to the limb of the Sun at specific position angles that
vary during the day \citep{1981SoPh...71..237C}. Especially in the
Winter, when the Sun passes lower in the sky, the celestial
object will undergo a significant rotation with respect to a slit
held at or near the parallactic angle. Since different types of 
observations may be more or less constrained by these considerations,
observation scheduling may have to take into account the different 
periods of the year when the parallactic angle changes more or less rapidly.

\begin{acknowledgements}
We thank the referee for useful comments that helped to improve the paper.
We are grateful for discussions and careful readings of the manuscript
by Gianna Cauzzi, Fabio Cavallini, and Alexandra Tritschler. 
The Mees meteorological data were kindly provided by Don Mickey. 
This work was supported by the Italian Research Ministry, PRIN-MIUR 2004.
\end{acknowledgements}

\newpage
%\Online
\appendix
\section{La Palma Calculations}
\label{appendixa}

We present here plots, similar to those shown in the main paper, calculated 
for Roque de los Muchachos, La Palma, Spain 
(latitude: 28.76\degree; longitude: -17.88\degree; altitude: 2350 m).
This is an alternate
site for the ATST and the location of other high-resolution solar telescopes 
such as the Swedish one-meter Solar Telescope (SST). These calculations will 
also apply to GREGOR which will be located on the island of Tenerife. For simplicity, 
we use the same meteorlogical conditions as for Halekal\=a. Due to the site's higher 
latitude, the behavior of the parallactic angle, for example, is different from 
that of a tropical site. 

\begin{figure*}
\centering
\includegraphics[bb=30 0 450 300,width=13cm,height=8.cm]{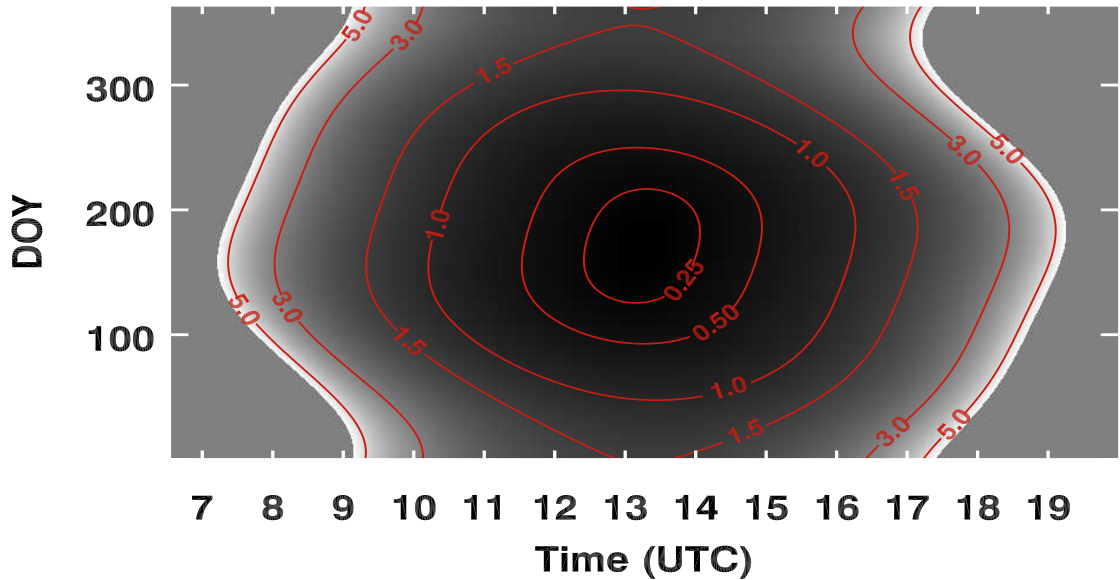}
\caption{Magnitude of the atmospheric
dispersion, in arcseconds, at La Palma for the Sun between
wavelengths of 400 and 850 nm. For each day of the year, shown on
the vertical axis, the dispersion is shown for all times when the
true elevation of the Sun is greater than 10\degree.}
\label{Figure:A1}
\end{figure*}

\begin{figure*}
\centering
\includegraphics[bb=30 0 450 300,width=13cm,height=8.cm]{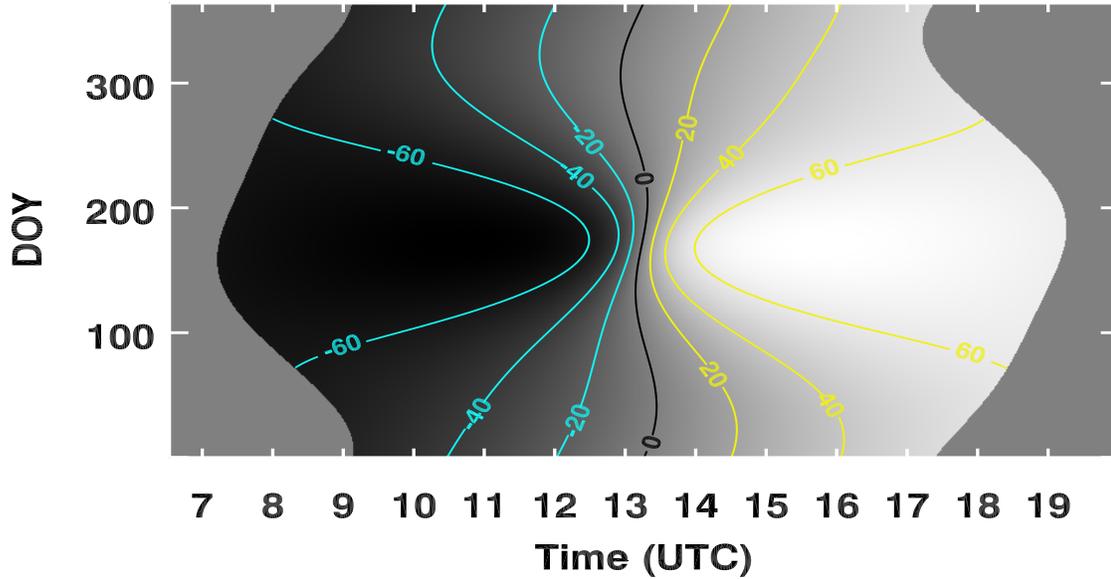}
\caption{The parallactic angle in degrees at La Palma for the Sun for 
all times when the true elevation of the Sun
is greater than 10 degrees. Negative angles are shown in blue, positive
angles in yellow.}
\label{Figure:A2}
\end{figure*}

\begin{figure*}
\centering
\includegraphics[bb=30 0 450 300,width=13cm,height=8.cm]{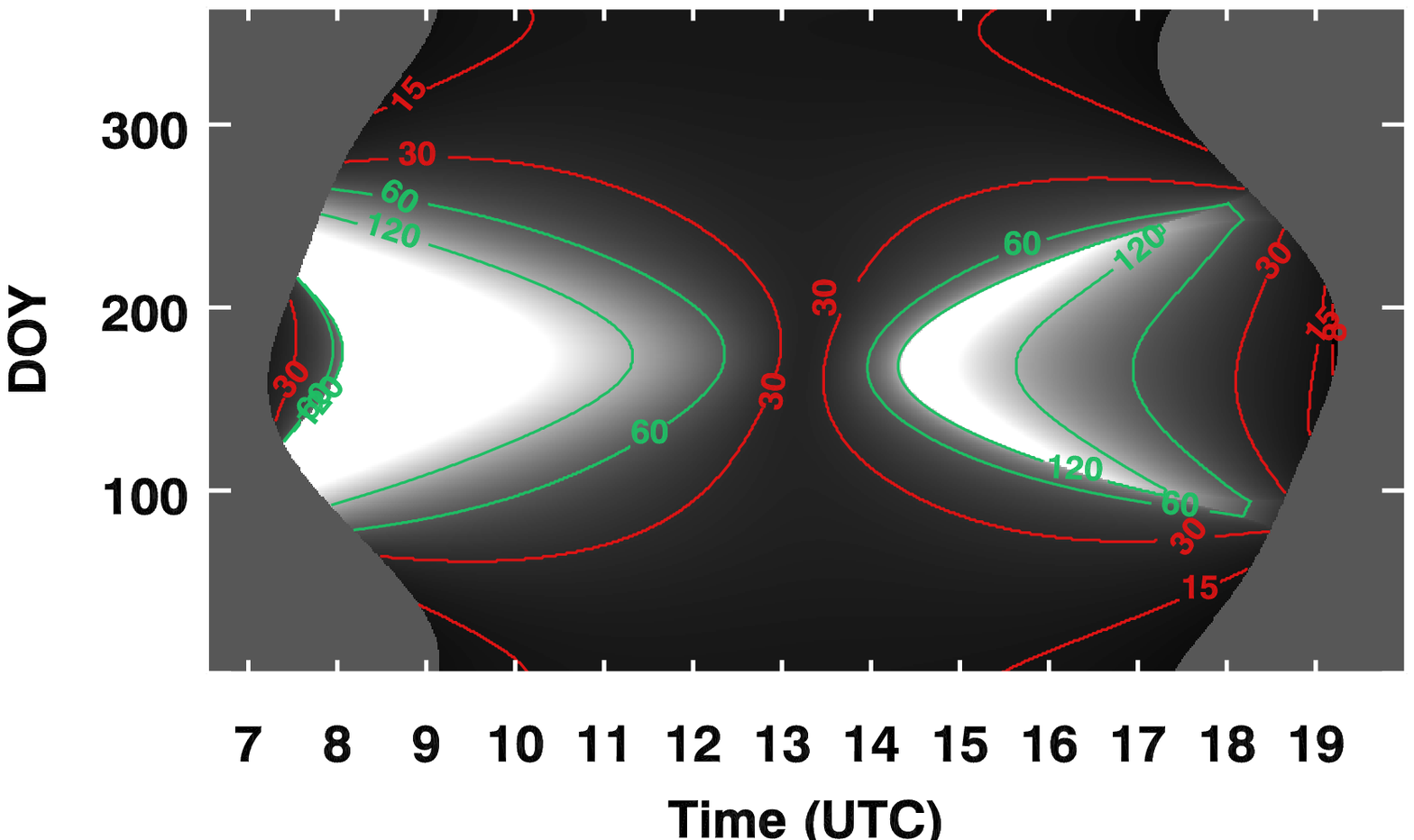}
\caption{The duration in minutes for which spectra
obtained at two different wavelengths will remain aligned within one slit
width. The slit is initially oriented perpendicular to 
the horizon but the image is rotated during the scan to maintain a constant 
orientation with respect to the celestial coordinates.
The figure is calculated for observations obtained at La Palma at 400 and 850
nm and with a slit width of 0.11\arcsec, a width typical of 
present observations \citep*[see {\it e.g.}][]{2005A&A...443L...7B}.}
\label{Figure:A3}
\end{figure*}
\end{article} 
\end{document}